# Fast Directed *q*-Analysis for Brain Graphs

Felix Windisch,[*] Florian Unger[†]

April 24, 2025

**Abstract**

Recent innovations in reconstructing large scale, full-precision, neuron-synapse-level connectomes demand subsequent improvements to graph analysis methods, to keep up with the growing complexity and size of the data. One such tool is the recently introduced *directed q-analysis*. We present numerous improvements, theoretical and applied, to this technique. On the theoretical side, we introduce modified definitions for key elements of directed *q*-analysis, which remedy a well-hidden and previously undetected bias. This also leads to new, beneficial perspectives to the associated computational challenges. Most importantly, we present a high-speed, publicly available, low-level implementation that provides speed-ups of several orders of magnitude on *C. Elegans*. Furthermore, the speed gains grow with the size of the considered graph. This is made possible due to the mathematical and algorithmic improvements as well as a carefully crafted implementation. These speed-ups enable, for the first time, the analysis of full-sized connectomes like those obtained by recent reconstructive methods.

Additionally, the speed-ups allow comparative analysis to corresponding null models, appropriately designed randomly structured artificial graphs that do not correspond to actual brains. This in turn, allows for assessing the efficacy and usefulness of directed *q*-analysis for studying the brain. We report on the results in this paper.

## 1 Introduction

Even a mere decade ago, structural neuroscience, namely the quest to explain brains extraordinary computational power through the *structure* of its network of neurons and synapses, was a field almost entirely devoid of biological data. The most prominent objects of study were *C. Elegans* and artificial, statistical reconstructions on the basis of select sparse biological facts like neuron-type dependent connectivity probabilities.

This changed enormously in recent years, as advancements in data processing allowed for scans of entire brain hemispheres of — compared to *C. Elegans* — terrifyingly intelligent fruit flies [4, 7].

Data alone is not sufficient, though. It must be analysed, which requires the development of tools, one of which we study here: *directed q-analysis.* In order to (even informally) motivate this particular approach, one must first introduce connectomic graphs as well as simplices, which form the building blocks. A mathematically rigorous exposition lies in Section 2.

*Connectomes:* The classical paradigm in neuroscience attributes most of the brain's computational power to neurons and synapses. Viewed from a network science perspective, a neuron is simply a point, and (chemical) synapses are directed connections from one point to another. Ignoring autapses, this results in a simple directed graph called the (neuron-synapse level) connectome.

*Simplices:* A certain motif or subgraph is both elementary and very prevalent: (directed) simplices. These are dense, acyclic subgraphs, where (ignoring direction) everything connects to everything, but which also feature a clear and unique direction of flow from a *source* neuron to a

---

[*]Technische Universität Graz, Austria, felix.windisch@tugraz.at, Corresponding Author, Inffeldgasse 16, 8010 Graz

[†]Technische Universität Graz, Austria, florian.unger@tugraz.at

*sink* neuron. The maximum simplex size ranges typically from 7 (*C. Elegans*) to 25 (*Drosophila*) neurons.

Studying simplices in isolation, or merely counting their occurrences would be a missed opportunity, as they form quite interesting structures. Natural questions to ask are:

- How do the flows from two distinct, but connected simplices interact?
- Are there long, robust highways in the graph, comprised of chains of simplices connected in specific ways?

**Directed $q$-Analysis** strives to answer these questions. It takes as input a connectomic graph and a triplet of integer parameters $(q, i, j)$, and outputs a directed graph where simplices become vertices and the presence of a directed edge indicates that the two incident simplices interact in the particular way described by $(q, i, j)$. Specifically, $q$ encodes the strength of the connection, and $(i, j)$ indicates where (with reference to the simplex flow of the two connected simplices, as it is found in the connectomic graph) the interacting section is located. Applied to all simplices in the original connectome, we obtain as the output $(q, i, j)$-digraph $\mathcal{Q}$ with simplices as nodes and $(q, i, j)$-relations between simplices as edges.

**Relation to Network Science.** The $(q, i, j)$-digraph $\mathcal{Q}$ then represents the network of simplices that interact in the $(i, j)$ direction. The focus on simplices is motivated by the observation that brain graphs contain an unexpected abundance of them [5], and they represent the logical extension of directedness to cliques. This network of simplices may exhibit drastically different properties to the original graph, but as it is also typically much larger, it is difficult to gain insight by direct inspection. Therefore, directed $q$-analysis in the narrow sense should not be seen as a replacement or variation type of network Aaalysis, but instead as a preprocessing step that enables deeper inspection under a certain, simplical light. To demonstrate, we showcase (in Section 4) how analysis of $\mathcal{Q}$ may yield novel insight that is not easily gained from the original graph.

While the number of shared vertices between simplices may be interpreted as the strength of their connection, one should not confuse them with edge weights: since both the input and $(q, i, j)$-digraph are unweighted, $q$ should instead be interpreted as filtering for a particular scale.

On first glance, the fully connected simplices used in $q$-analysis appear similar to the strongly connected communities from classic community detection algorithms. There are however major differences between these two: While community detection only aims to segment the graph into (mathematically rather loosely defined) communities, $q$-analysis instead searches for simplices (a very precisely defined subgraph) in the graph and analyses their interaction. Any vertex may be part of arbitrarily many simplices, but by definition, only in one community.

## 1.1 Contributions

**Mathematical.** We propose a novel definition of directed $q$-nearness that fixes a previously undiscovered hidden bias while remaining equivalent in important edge-cases and near-equal everywhere else.
**Algorithmic.** We introduce new algorithms that significantly accelerate the computation of the novel and original definition, enabling application on larger networks that was previously infeasible.
**Implementation and Accessibility.** We provide a heavily optimized open-source `Rust` implementation for a suite of $q$-analysis algorithms. They include new features and exhibit major gains in accessibility owing to a GUI application and `python` bindings. In benchmarks, our implementation outperforms the previous state of the art by a factor of at least $10^6$ on large graphs thanks to optimizations both mathematical and in the implementation. This factor only becomes bigger as the graphs grow, since the new algorithm lies in another asymptotic complexity class.
**Guidelines for Application.** We propose and demonstrate that constraining maximal and minimal simplex dimension refine the results and lead to easier interpretability.

**Comparison to Null Models.** As a consequence of this newfound efficiency, we demonstrate the usefulness of $q$-analysis techniques by analysing connectomes at a finer level than previously possible. We also assess $q$-analysis' ability to detect features in connectomic graphs by comparing them to closely matching null models.

## 1.2 Related Work

In [6], Riihimäki extended the ideas of undirected $q$-analysis, which were previously developed in the 1970s by Atkin (see e.g. [1]), to the directed version extended upon in this work.

Directed $q$-analysis can be understood to be part of the greater field of Topological Data Analysis, being a hybrid of topological and network approaches. Reimann et al. [5] were the first to take the topological viewpoint in the context of connectomes, or even to just extend the notion of clique complexes to directed flag complexes. Further important contributions, especially on the computational side, were made in [2]; our work and implementation were much inspired by the algorithmic ideas introduced there.

The more theoretical aspects of this, in particular a formal description, proof of correctness and computational complexity analysis may be found in [9].

Section 2, 3.2, 3.3, 3.4 and Appendix A of this work present a highly condensed version of a 2024 Master Thesis by one of the authors [11]. More exhaustive details can be found there, as well as a more historically motivated way to introduce directed $q$-analysis and many alternative algorithmic approaches.

## 2 Theory

In this section, we formally introduce the definitions required for directed $q$-analysis.

**Definition 2.1** (Graphs, Simplices).

- A *simple directed graph* is a pair $G = (V, E)$ of a finite set of vertices $V$ and a relation $E \subseteq (V \times V) \backslash d_V$, where $d_V = \{(v, v) \mid v \in V\}$.

- A $d$-dimensional (directed) *simplex* is an acyclic clique, i.e. a subset $(v_0 v_1 ... v_d)$ of $V$ such that $(v_i, v_j) \in E \ \forall i < j$. Note that there is a strict order on the vertices in a simplex.

- We call $\mu_\sigma$ a *face* of $\sigma$ if its vertices are a subset of the vertices of $\sigma$ and the vertices occur in the same order. We write $\mu_\sigma \hookrightarrow \sigma$ or $\sigma$ *includes* $\mu_\sigma$.

In undirected graphs, edges denote mutual connection and cliques are sets of vertices that share a strong mutual connection. In a directed graph, edges define both connection *and* direction, thus a simplex should constitute a set of vertices that are both connected and define a unique (cycle-free) direction.

One potentially surprising side-effect is the possibility to have multiple simplices over the same set of vertices. Indeed, $n$ vertices with bidirectional edges between each other contain one clique, but up to $n!$ simplices: one for each possible order on the vertices.

**Definition 2.2** (Flag Complexes). Let $G = (V, E)$ be a simple directed graph and $D$ the dimension of the highest-dimensional simplex in $G$. The *directed flag complex* $\Sigma$ of $G$ is a tuple $(\Sigma_0, \Sigma_1, ..., \Sigma_D)$, where $\Sigma_d$ is the set of all $d$-dimensional simplices in the graph. Observe that $\Sigma_0 = V$ and $\Sigma_1 = E$. We write $\Sigma_{>q}$ to denote the subset $(\Sigma_q, \Sigma_{q+1}, ..., \Sigma_D)$.

**Definition 2.3** (Face Maps). Let $\sigma \in \Sigma_n$. The $i$th *face map* $\hat{d}_i(\sigma)$ maps $\sigma$ to its $i$th face, which is obtained by removing the $i$th vertex from $\sigma$. If $i > n$, the last vertex of $\sigma$ is removed. Whenever $i \leq n$ can be guaranteed, we omit the ˆ and write $d_i(\sigma)$ instead. If the size of the simplex is not explicitly given, we use $d_\infty$ to denote the face map that always removes the last vertex.

**Example 2.4.** Consider the following directed graph:

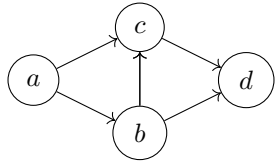

The associated flag complex is described by $\Sigma = (\Sigma_0, \Sigma_1, \Sigma_2)$, where $\Sigma_0 = V = \{(a),(b),(c),(d)\}$, $\Sigma_1 = E = \{(ab),(bc),(ac),(cd),(bd)\}$ and $\Sigma_2 = \{(abc),(bcd)\}$.

The three faces of $(abc)$ are $\hat{d}_0(abc) = (bc)$, $\hat{d}_1(abc) = (ac)$ and $\hat{d}_2(abc) = (ab)$ (note that $\hat{d}_2(abc) = \hat{d}_i(abc)$ for $i > 2$).

The directed generalisation of $q$-connectedness was first explored in [6] with the following definition:

**Definition 2.5** ($(\widehat{q,i,j})$-nearness)**.** Let $\Sigma$ be a directed flag complex and $(\sigma, \tau)$ be an ordered pair of simplices $\sigma, \tau \in \Sigma_{>q}$. Let $(\hat{d}_i, \hat{d}_j)$ be an ordered pair of face maps. Then $\sigma$ is $(\widehat{q,i,j})$-near to $\tau$ if either of the following conditions is true:

[I] $\sigma \hookrightarrow \tau$,

[II] $\hat{d}_i(\sigma) \hookleftarrow \alpha \hookrightarrow \hat{d}_j(\tau)$ for some $\alpha \in \Sigma_q$.

We introduce a novel, slightly modified version of this definition. The difference, as also depicted in Figure 1 is in where the face maps are applied — for the original one, right at the top at the level of $\sigma$ and $\tau$, for the novel one it is always applied to faces of dimension $q+1$.

**Definition 2.6** ($(q,i,j)$-nearness)**.** Let $\Sigma$ be a directed flag complex and $(\sigma, \tau)$ be an ordered pair of simplices $\sigma, \tau \in \Sigma_{>q}$. Let $(d_i, d_j)$ be an ordered pair of face maps with $i, j \in \{0, ..., q+1\}$. Then $\sigma$ is $(q,i,j)$-near to $\tau$ if either of the following conditions is true:

[I] $\sigma \hookrightarrow \tau$,

[II] There exist a $q$-simplex $\alpha \in \Sigma_q$ and two $(q{+}1)$-simplices $\mu_\sigma \hookrightarrow \sigma$ and $\mu_\tau \hookrightarrow \tau$ such that $d_i(\mu_\sigma) = \alpha = d_j(\sigma_\tau)$.

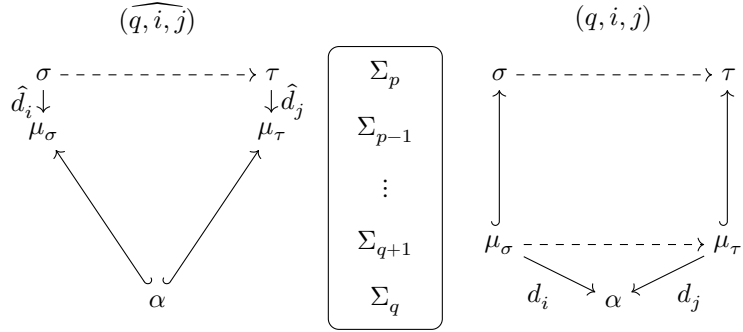

Figure 1: Condition for two simplices $\sigma$, $\tau$ to be $(\widehat{q,i,j})$-near (left) or $(q,i,j)$-near (right) by criterion [II] with dimension indicator in the middle. Recall that $\hookrightarrow$ denotes inclusion and $\dashrightarrow$ denotes $(q,i,j)$-nearness.

While these two definitions presented are subtly different, they both agree either in the $(0, \infty)$-direction and when both $\sigma$ and $\tau$ are of dimension $q+1$. While there are examples where $\sigma$ and $\tau$ are $(q,i,j)$-near but not $(q,i,j)$-near and vice versa, where neither of these definitions is more strict or more lenient, they simply differ in which direction they attribute $q$-nearness to.

**Proposition 2.7.** *1. Two simplices $\sigma, \tau$ are $(q, 0, \infty)$-near iff they are $(\widehat{q, 0, \infty})$-near. The same holds true for the directions $(0,0)$, $(\infty, \infty)$ and $(\infty, 0)$.*

*2. If $\sigma, \tau \in \Sigma_{q+1}$ are $(q,i,j)$-near, they are also $(\widehat{q,i,j})$-near and vice versa.*

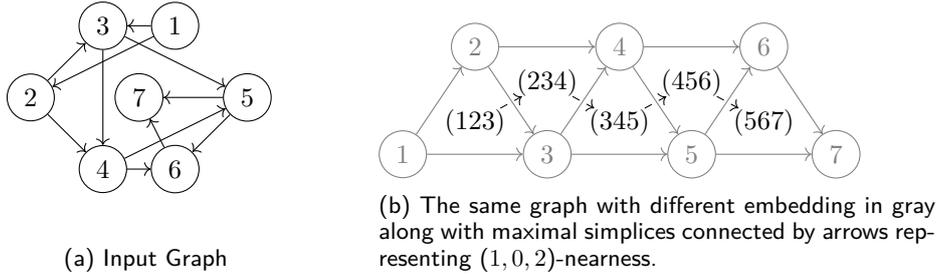

(a) Input Graph

(b) The same graph with different embedding in gray along with maximal simplices connected by arrows representing $(1, 0, 2)$-nearness.

Figure 2: Important structures of graphs become apparent in the $(q, i, j)$-digraph.

3. If two simplices $\sigma$ and $\tau$ are $(q, i, j)$-near, they are also $\widehat{(q, k, l)}$-near for some $k, l \in \mathbb{N}$ and vice versa.

A proof of Proposition 2.7 and a much more exhaustive exploration of the similarities and slight differences of the definitions can be found in [11] or [9].

As one is usually not only interested in the $(q, i, j)$-nearness of two specific simplices, but rather the set of all $(q, i, j)$-nearness relations, one defines the $(q, i, j)$-digraph which encompasses all of them:

**Definition 2.8** (($(q, i, j)$-Digraphs). Let $G$ be a simple, directed graph with directed flag complex $\Sigma$. The $(q, i, j)$-digraph $\mathcal{Q} = \{\Sigma_{\geq q}, E^{\mathcal{Q}}\}$ of $G$ contains an edge $(\sigma, \tau)$ iff $\sigma \in \Sigma_{\geq q}$ is $(q, i, j)$-near to $\tau \in \Sigma_{\geq q}$. Analogously, $\hat{\mathcal{Q}} = \{\Sigma_{\geq q}, E^{\hat{\mathcal{Q}}}\}$ contains connections between $\widehat{(q, i, j)}$-near simplices.

As $q$ represents the size of the shared simplex, increased values lead to a sparser $(q, i, j)$-digraph, where only high-dimensional simplices with very strong connections are included. The values of $i$ and $j$ determine the type of connection that is included in the $(q, i, j)$-digraph: In the extreme case $i = 0$ and $j = \infty$, the $(q, i, j)$-digraph will contain connections from simplices to any other simplex that continues its direction (e.g. if one simplices' sink is the source of the other), this is illustrated in Figure 2. Conversely, more similar values for $i$ and $j$ would lead to connections of simplices whose edge direction runs in parallel.

## 3 Proposed Enhancements to Directed $q$-Analysis

### 3.1 Bottom-Up Definition

We see the novel definition of $(q, i, j)$-nearness as an improvement over the original one on two accounts:

**Avoiding $\hat{d}$.** With the original definition, the only reasonable upper bound of $i, j$ is the maximal simplex dimension of the graph. The face maps are applied to all $\sigma, \tau$ of potentially smaller dimension, which raises the need to introduce $\hat{d}$. This cut-off seems like a harmless technicality at first, but introduces a bias on the indices. If they are clamped to the range of $[0, q + 1]$, later faces of higher-dimensional simplices in the graph will not contribute to the analysis. Whereas without limiting the range, connections will be over-counted for high indices.

We admit that finding a better normalization method is tricky, but luckily, the bottom-up definition introduced here simply does not require normalization at all: As the face maps are only ever applied to $(q+1)$-simplices, the values are naturally bounded: $0 \leq i, j \leq q + 1$. Simultaneously this enables picking reasonable ranges for valid $i, j$ in cases where the maximal simplex dimension is not know a priori.

**Upward Closure.** Another neat feature of the novel definition is the property of upward-closure: If $\sigma, \tau$ are $(q, i, j)$-near, then every simplex containing $\sigma$ is also $(q, i, j)$-near to every simplex containing $\tau$. This property not only helps with theoretical reasoning, but also reduces the computational effort needed to calculate the $(q, i, j)$-digraph.

## 3.2 Algorithmic Improvements

As the problem of computing directed flag complexes from graphs has been thoroughly explored in [2] and [5], we consider the flag complex as given, briefly describing in the following only the computations onwards. A full, more formal description of the algorithms together with an analysis of correctness and computational complexity may be found in [9]. We also describe the (pre-existing) top-down approach only for $\widehat{(q,i,j)}$-nearness and the novel hybrid approach only for $(q,i,j)$-nearness as these definitions inspire the respective approaches. Consult [9] for the other combinations.

**Top-Down Approach.** Computing whether two simplices $\sigma, \tau$ are $\widehat{(q,i,j)}$-near is algorithmically straightforward:

Removing the $i$th index from $\sigma$ and the $j$th from $\tau$ are both trivial operations and checking for an order-preserving subset $\alpha$ is computationally also not demanding.

We call this approach the top-bottom approach and it is perfectly valid (and the best) approach to check $\widehat{(q,i,j)}$-nearness for two single simplices.

By iterating this procedure over all pairs of $\sigma, \tau$, the $\hat{\mathcal{Q}}$-digraph can be computed in a very simple manner. This is exactly the approach used in [6]. Generously assuming that calculating a single check is done in constant time, this approach has an asymptotic runtime of $\mathcal{O}(|\Sigma_q|^2)$, the size of the adjacency matrix of $\hat{\mathcal{Q}}$.

The problem lies in the extreme sparsity of $\hat{\mathcal{Q}}$: For any naturally occurring connectome, most simplices don't even share a single vertex! Yet, we have to perform checks for every pair of them.

**Hybrid Approach.** In contrast, instead of checking every simplex pair as with the Top-Down approach, we start at the shared $q$-simplex $\alpha$. We compute the set of all $\mu_\sigma, \mu_\tau \in \Sigma_{q+1}$ such that $d_i(\mu_\sigma) = d_j(\mu_\tau) = \alpha$ for all $\alpha \in \Sigma_q$ and then find all pairs of simplices $\sigma, \tau$ that include $\mu_\sigma, \mu_\tau$. These, as well as their supersimplices up to dimension $q$, are then, by the principle of upward closure, $(q,i,j)$-near.

This approach has a catch, though: finding $\mu_\sigma, \mu_\tau, \sigma, \tau$ would be very costly if starting from $\alpha \in \Sigma_q$, as we would have to find these higher-dimensional simplices by adding the correct vertices to $\alpha$, for which there is only a brute-forte search across all vertices of the graph. To avoid this cost, we first precompute all inclusion relations $\hookrightarrow$ in the flag complex (which are already required due to condition [I]) in a top down pass and during that also cache the sets of potential $\mu_\sigma, \mu_\tau$. This is done from the highest dimension of the supplied flag complex downwards. As removing vertices from simplices is a simple $\mathcal{O}(1)$-operation instead of the $\mathcal{O}(|V|)$-cost of *adding* vertices, this provides a significant speedup.

This hybrid approach (as it incorporates top-down strategies ans well as bottom-up strategies) is ideal for enumerating *all* $(q,i,j)$-nearness relations, as it never has iterate over any pairs that are not $(q,i,j)$-near. Indeed, under mild assumptions, it is asymptotically time-optimal as stated in [9].

## 3.3 Implementation and Accessibility

We improve accessibility of *q*-analysis by providing a high-performance, highly parallelized, memory-friendly implementation of both top-down and hybrid algorithms, which are, both for $\widehat{(q,i,j)}$-nearness and $(q,i,j)$-nearness, available at `https://github.com/FelixWindisch/DirQ`. Binary executables bundled with documentation of the GUI are available at `https://repository.tugraz.at/records/2dmte-zxw28`. The software package is written in `Rust`, offering additional speed-ups like other fully compiled languages like `C/C++`, while `python`-bindings allow users to reap all the benefits of higher-level languages. For `x86-64` linux and windows, the package can be comfortably installed using PyPI with the command `pip install directed_q`.

We demonstrate the efficiency of the hybrid algorithm using benchmarks performed on an AMD EPYC 7543 32-Core processor and 512 GB of DDR4 RAM. The resulting computation times on various test graphs can be seen in Table 1. We observe up to 300.000x speedup of for bigger networks like BBP[3] and consistently faster times for the novel definition.

|  |  |  |  | Top-Down | | Hybrid (ours) | |
| --- | --- | --- | --- | --- | --- | --- | --- |
| Graph | Nodes | Edges | $q$ | $\widehat{(q,i,j)}$ | $(q,i,j)$ | $\widehat{(q,i,j)}$ | $(q,i,j)$ |
| *C. Elegans* [10] | 279 | 2194 | 3 | 10.68s | 11.16s | 27.93ms | 25.72ms |
| Erdős–Rényi | 1000 | 50k | 3 | 16.98s | 17.34s | 765.64µs | 455.97µs |
| BBP [3] | 31k | 7.6M | 4 | 2062.22s | 2054.21s | 14.60ms | 5.57ms |
| BBP [3] | 31k | 7.6M | 3 | > 48h | > 48h | 45.85s | 36.18s |

Table 1: Benchmark results for the `Rust` implementation of the hybrid and top-down algorithm. Top-down runtimes were measured using our own optimised implementation, already 10x faster than the original `python` implementation in [6].

## 3.4 Best Practices: Limiting Simplex Dimension

By limiting the minimal and maximal simplex dimension in the $q$-digraph we not only save on computation time, but also avoid problems in the definition of directed $q$-nearness.

**Upper Limit.** In a theoretical investigation of directed $q$-analysis we found adverse effects to including simplices with dimension higher than twice the value of $q$.

A disproportionately large simplex can act as a kind of "hub" that can connect, through the inclusion property, smaller simplices even against its own direction. This allows small simplices to "jump" against the $(i,j)$-direction of the flow, severely hindering interpretability. We thus recommend a cut-off of the flag complex dimension at some value between $q+1 \leq D_{\text{MAX}} \leq 2q$.

A detailed, thorough example (with figure) of this unintuitive effect is given in Appendix A.

**Lower Limit.** Conversely, we also propose to exclude all simplices of lowest dimension $q$ from the $q$-digraph. If any simplex $\sigma \in \Sigma_q$ is $(q,i,j)$-near to any simplex $\tau$, then $\sigma \hookrightarrow \tau$. Therefore, the more interesting condition 2 involving face maps never applies to $q$-dimensional simplices. Additionally, this "bottom layer" of inclusion connections is identical for all values of $i,j$ and leads to certain graph properties (such as the number of connected components) to be identical for all $i,j$.

## 4 Applying and Evaluating $q$-Analysis Techniques

It is often difficult to tell if particular results from directed $q$-analysis are simply an ingrained part of the method or significant structural differences. Thus, in order to test hypotheses empirically, we need to compare the results on connectomes to a set of null models. Standard null models used in network science such as Erdős–Rényi graphs or configuration models do not make sense in the context of directed $q$-analysis, as they would lead to drastically reduced numbers of simplices. As we are interested in the interaction of the simplices, our null model should contain the same amount and dimension of simplices as the original graph, but shuffle their configuration.

We test the proposed improvements on the reconstruction of *C. Elegans* (in the version found in [10]) and layer 1-4 of the stochastic reconstruction of a somatosensory cortex of a rodent (hereby referred to as BBP, see [3]) These graphs were chosen since they are the only ones for which these strict null models are available: The authors of [8] provide roughly 300 of them for both *C. Elegans* and BBP.

In each experiment we compute, for each pair of $(i,j)$, the $(q,i,j)$-digraph of the brain graph in question (either *C. Elegans* or BBP) as well as the $(q,i,j)$-digraphs for each of these 300 graphs of the null model associated to that brain graph. These $(q,i,j)$-digraphs (both original and null model) are then, each by themself, analysed by means of various graph metrics (degree centrality, closeness centralisation, approximated longest path and number of connected components). We then report and interpret the $z$-scores for each experiment and each pair of $(i,j)$. This measures the number of standard deviations the brain graph differs from the mean of the null model.

**Degree Centrality.** When examining strong, redundant connections in connectome networks, one interesting question to ask is if there are certain central simplices that act as important hubs for connections. One measure for the importance of a node is *degree centrality*, which is defined as the ratio of connected nodes to the total number of nodes in the graph. A graph that contains singular nodes of very high importance would thus have an increased *maximal degree centrality*. Indeed, as can be seen in Figure 3, for $i = 0$ and $j = q + 1$ (and vice versa), we find that the *maximal degree centrality* of the *C. Elegans* graph exceeds the mean of the null models by 5.6 standard deviations. This is particularly enlightening, as the directions $(0, q + 1)$ and $(q + 1, 0)$ correspond to forward-directed flow (compare Figure 2). We find that this is a result of particularly high in-/out-degree centrality for minimal and maximal values of $i$ and $j$ respectively.

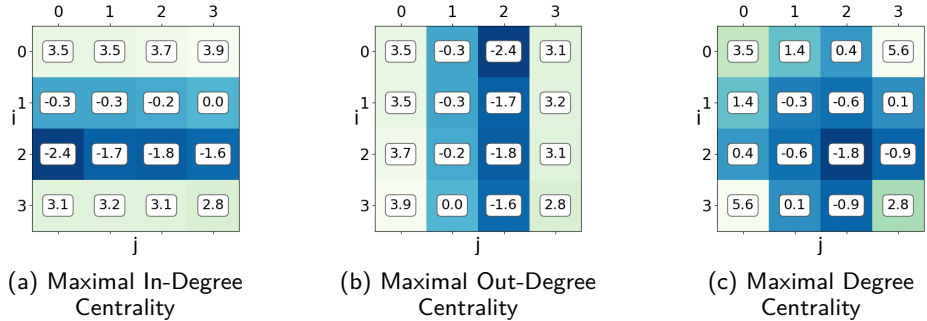

(a) Maximal In-Degree Centrality

(b) Maximal Out-Degree Centrality

(c) Maximal Degree Centrality

Figure 3: Resulting $z$-scores (from the mean of null models) of various degree centralities of the $(2, i, j)$-digraph for *C. Elegans* with $q = 2$, $D_{\text{MAX}} = 3$ for varying values of $i$ and $j$. High $z$-scores indicates the presence of well-connected *hubs* featuring in the brain graph, but not the null model.

**Closeness Centralization.** Closeness centrality measures the importance of a vertex not only among its neighbours, but considers the whole graph. Each vertex $i$ has a centrality score $c_i$ which is the reciprocal of the sum of the length of the shortest directed paths between that vertex and all other vertices.[1] With $c_{\text{max}}$ being the maximal centrality value, the *closeness centralization* of the graph is is defined as $\sum_{i=1}^{N}(c_{\text{max}} - c_i)$. High closeness centralization indicates the existence of a (small) collection of vertices being very central and the directedness implies that these vertices behave more like sources. One the other hand, if we were to observe high centralization in the flipped graph (i.e. all edges are flipped around or the adjacency matrix being transposed), it would indicate the existence of some kind of sink.

Again, we investigate the distance of *C. Elegans* compared to its null model. The null model exhibits centralization values quite similar for both the $(0, \infty)$- and $(\infty, 0)$-direction. In comparison, for *C. Elegans* we find that with $(q, i, j) = (4, 0, \infty)$, the closeness centralization is 3.8 standard deviations *higher* than the mean, whereas for the opposite direction, $(q, i, j) = (4, \infty, 0)$, *C. Elegans* exhibits a closeness centralization 1.4 standard deviations *lower*.

As flipping all edges in the $(q, i, j)$-graph results exactly in the $(q, j, i)$-graph (see [9]), this finding could be interpreted as the existence of some kind of control hub simplices, while simultaneously indicating a lack of central collecting or receiving simplices.

**Longest Path.** The initial question on the presence of long, robust highways was already discussed in [6], where the $(q, i, j)$-digraphs were analysed using an approximation of the longest path length. While this analysis was conducted at a coarse level of $q = 4$, the efficiency gains from the new algorithm allow us to conduct this same experiment at a much finer scale of $q = 2$. We also omit $q$-level simplices as recommended in Section 3.4. The results in Figure 4 show even more pronounced effects than the original work. We also find that the extreme $(0, 3)$ and $(3, 0)$ directions exceed the mean of the null models by 6.5 and 7.6 standard deviations respectively (not shown).

---

[1] More precisely, to handle pairs of nodes without a directed path between them, we used the Wasserman and Faust method as described here: https://networkx.org/documentation/networkx-3.4.2/reference/algorithms/generated/networkx.algorithms.centrality.closeness_centrality.html

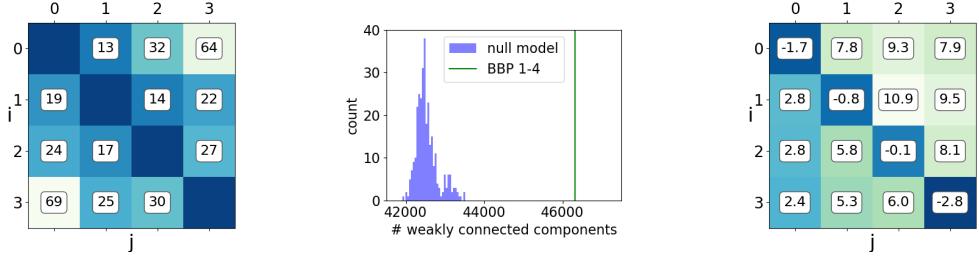

Figure 4: **Left:** Approximated longest path lengths on the $(2, i, j)$-digraph of *C. Elegans* with $D_{\mathsf{MAX}} = 4$. **Middle:** Histogram of the number of weakly connected components on the $(3, 0, 4)$-digraph of BBP layer 1-4 with $D_{\mathsf{MAX}} = 6$. **Right:** $z$-scores of number of strongly connected components on the $(2, i, j)$-digraph of *C. Elegans* with $D_{\mathsf{MAX}} = 5$.

**Connected Components.** An obvious question that presents itself when working with $(q, i, j)$-digraphs is how well-connected they are. This can be described by the number of *weakly* and *strongly connected components*. All nodes within the same weakly connected component are reachable by at least one node of that component, whereas in a strongly connected component *all* nodes fulfil this criteria. A higher number of connected components implies that the graph is better connected locally than globally. Indeed, in Figure 4 we observe a significant ($z$-score =13.5) increase in the number of weakly connected components of the BBP graph as opposed to the null model. This pattern, but for strongly connected components repeats in *C. Elegans*, in particular for high values of $j$.

**Efficacy of Directed $q$-Analysis.** These experiments prove that the analysis of $(q, i, j)$-digraphs is able to distinguish brain graphs from very close null models. Importantly, the $(q, i, j)$-digraph may also uncover structure that would not be immediately apparent in the original graph: Inspecting closeness centralization we observe that the $(4, 0, \infty)$-digraph of *C. Elegans* has a $z$-score of 3.8, whereas the original graph has a $z$-score of 0.6, which is non-significant. In certain cases, the structure on the simplices differs drastically to the structure on the vertices in the original graph: Whereas the connectome of *C. Elegans* has a *negative* maximal degree centrality $z$-score of $-3.4$, the $z$-score on the $(2, 0, \infty)$-digraph of the connectome is *positive*, namely 5.6.

## 5 Summary and Outlook

**Summary.** We believe that our efforts help the previously borderline esoteric method of $q$-analysis to become more practical and useable. With the demonstration that directed $q$-analysis is able to distinguish between biological graphs and null models (by, in some cases, more than 13 standard deviations), proof that it is able to easily discern brain graphs even when the null-modell compared to is extremely close. This leads to the conclusion that directed $q$-analysis *can* be used and *should* be used if interest lies in analysing simple, directed graphs.

**Discussion.** Directed $q$-analysis can show interesting results that have solid interpretations as presented in Section 4. Clearly, it does have capability to explain some aspects of the simplicial structure of these networks. At the same time, one has to be very careful about the choice of parameters, as choosing values that are even slightly off may lead to degenerate outputs like empty graphs or outputs that exceed reasonable storage capacity.

Additionally, $q$-analysis is only applicable where flag complexes exhibit sufficient simplicial structure, but computing the entire flag complex is still feasible.

Despite significant time invested, some results such as spikes at particular parameter settings leave us occasionally puzzled. The analysis is picking up on something, but due to the complexity of $q$-analysis the root cause is often difficult to grasp.

**Outlook.** While sufficient for all current datasets, future datasets might require revisiting the algorithmic approaches for more parallelized approaches. Some potential approaches are already discussed in [11]. Another cause to revisit is the natural extension of $q$-analysis with a filtering approach, filtering over synapse weights for more refined insights.

A further investigation of connectomes, one which respects the existence of inhibitory and excitatory neurons, is currently restricted to analysing the respective homogenous subnetworks. This limits studying interaction between these two groups of neurons though, immediately warranting a signed extension of directed $q$-analysis. Unfortunately, this requires revisiting and extending even the most fundamental mathematical building blocks.

*Acknowledgements.* We would like to express our gratitude for their time, ideas and general help (alphabetic order): Olga Diamanti, Michael Kerber, Robert Legenstein, Henri Riihimäki, Markus Steinberger.

*Competing Interest Statement.* On behalf of all authors, the corresponding author states that there is no conflict of interest.

*Funding.* This research did not receive any specific grant from funding agencies in the public, commercial, or not-for-profit sectors.

*Author Contribution.* Conceptualisation: FU. Data Curation: FU. Formal Analysis: FW, FU. Investigation: FW. Software: FW. Resources: FU, FW. Visualisation: FW. Writing - original draft: FW, FU. Writing - review & editing: FU, FW.

## A  Clipping Maximal Simplex Dimension

As already briefly discussed in Section 3.4, not clipping maximal simplex dimension can lead to unintuitive results.

One such case is depicted in the following graph:

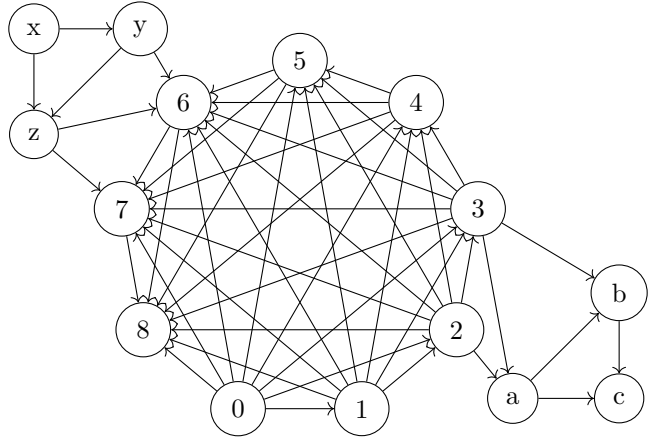

The graph is composed of two smaller simplices that are connected on either sides to a large, central simplex. The direction of flow of the central simplex points from the intersection with $(abc)$ towards the intersection with simplex $(xyz)$. As seen in previous examples, the direction $i = 0, j = \infty$ is typically characterized by continued forward flow. Unintuitively, and for both definitions, the $(1, 0, \infty)$-digraph in this example contains the following path from $(xyz)$ to $(abc)$, seemingly traversing the central simplex against the direction of flow:

$$(xyz) \dashrightarrow (yz6) \dashrightarrow (z67) \dashrightarrow (678) \hookrightarrow (12345678) \dashrightarrow (23a) \dashrightarrow (3ab) \dashrightarrow (abc)$$

The surprising result originates specifically from the inclusion edge $(678) \hookrightarrow (12345678)$. As criterion [I] is not concerned with the direction of simplices, it enables paths that traverse against the direction of the underlying graph through sufficiently large simplices.